\documentstyle[prb,aps,twocolumn]{revtex}

\begin{document}
\title{Anisotropic spin-glass-like  and quasi-one-dimensional magnetic behaviour in an 
intermetallic compound, Tb$_2$PdSi$_3$}
\author{P.L. Paulose and E.V. Sampathkumaran}
\address{Tata Institute of Fundamental Research, Homi Bhabha Road, Colaba,
Mumbai - 400005, India}
\author{H. Bitterlich, G. Behr and W. L\"oser}
\address{Leibniz Institute for Solid State and Materials Research Dresden,  Postfach 270016, 
D-01171 Dresden, Germany.}
\maketitle

\begin{abstract}
We report temperature dependent ac susceptibility ($\chi$) measurements on a high quality single crystal of Tb$_2$PdSi$_3$, crystallizing in a AlB$_2$-derived hexagonal structure. This compound is found to exhibit features attributable to quasi-low-dimensional magnetism at high temperatures and {\it anisotropic} spin-glass-like behavior  at low temperatures ($<$ 10 K) with an unusually large frequency dependence of peak temperature in ac $\chi$(T). This compound thus presents a novel situation in {\it metallic magnetism}, considering that the former  phenomenon is normally encountered only among insulators, whereas the anisotropic spin-glass-like behavior, to our knowledge,  has not been known in stoichiometric intermetallic compounds.     
\end{abstract}
\vskip3cm

\maketitle

The field of magnetism continues to be an exciting one due to the discovery of magnetic materials with unusual characteristics leading to the discovery of  novel phenomena.  In recent years, there is a tremendous increase of  activities on the materials with  magnetic frustration effects leading to spin-glass freezing\cite{1} on one hand, and with low-dimensional effects\cite{2} on the other hand. As far as spin-glass freezing is concerned, it has been generally assumed that this phenomenon is isotropic in stoichiometric compounds.    With respect to low-dimensional magnetism,    
the focus has always been on insulators in which the magnetic ordering is mediated by superexchange mechanism via the intervening non-magnetic atoms. However, the existence 
of  low-dimensional effects, to our knowledge, has not been  
demonstrated among {\it intermetallic} compounds. The existence of a compound with these properties in a metallic environment would be  surprising, as the indirect exchange interaction via the conduction electrons in metals, being of a long-range type, is expected to destroy low-dimensional character in favor of three dimensional interaction.  
In this article, we bring out that the {\it intermetallic} compound, 
Tb$_2$PdSi$_3$,  exhibits certain unusual strongly anisotropic features in the temperature (T) dependent 
ac susceptibility ($\chi$), both above and below long range magnetic ordering temperature (close to 24 K), which appear to have some bearing to the above assumptions.

This compound crystalizes in a AlB$_2$-derived hexagonal structure.\cite{3} It is worthwhile stating that the Pd-based ternary compounds R$_2$PdSi$_3$ (R= Rare-earth) have been 
found to exhibit many interesting and unusual magnetic and transport 
properties.\cite{4,5,6,7,8,9,10,11,12,13,14,15,16} The properties of the Tb compound are particularly notable.  While the reader may see references 6 (on polycrystal) and 12 (on single 
crystal) for many other interesting properties of this compound, we recall here only 
those which are relevant to the present article.   
All the results conclusively establish that long range magnetic 
ordering sets in below about 24 K. However, the nature of the magnetism appears to be very 
complex with a strong evidence for the existence of ferromagnetic 
correlations along basal plane and antiferromagnetic  interplane interaction 
(along c-direction); the moment orientation appears to be on the basal plane.   
In addition, the dc $\chi$ data reveal one more  
transition  below about 10 K in the polycrystal,\cite{6} 
whereas the data for the single crystal reveal\cite{12} more transitions at 
lower temperatures depending on the orientation and the (magnetic field and thermal cycling)  
history of the sample.  Another puzzling observation in  the dc $\chi$(T) plot  is that there is 
a peak  at about 55 K for H//[0001] (where H is the magnetic field). This peak, extending 
over a wide temperature range, is very broad, reminiscent of that predicted 
for one-dimensional magnetism.\cite{17}  

The single crystalline sample employed in the present investigation is the 
same as that in Ref. 12. The ac $\chi$ measurements were performed in the T 
interval 2 - 150 K at various frequencies ($\nu$= 1.2, 12, 120 and 1220 Hz) 
for two crystallographic orientations [10$\bar1$0] and [0001] 
 with respect to the direction of the ac field (2 Oe) {\it on the same 
piece} employing a 
commercial (Quantum Design) superconducting quantum interference device.

The results of ac $\chi$ measurements are shown in Figs. 1 and 2 for the two 
orientations. It is distinctly clear from the  figure 1 that there is a 
peak at 23.6 K in the real part ($\chi$$\prime$) of ac $\chi$ and the 
peak temperature is  independent of $\nu$.   This establishes that the 
compound undergoes long range magnetic ordering, but not spin-glass, at 23.6 K. 
The corresponding feature (appearing as a kink at the same temperature) is extremely weak  
for H//[0001].   There is also a  
peak in the imaginary part ($\chi$$\prime\prime$) for 
[10$\bar1$0] at the same temperature, but considering that the peak position 
is $\nu$-independent,  one must conclude\cite{1} that the anomalies at 23.6 K do not originate from 
spin-glass freezing. 

{\it The first observation  of emphasis} is that, 
as the T is lowered, one observes another peak around 10 K for [10$\bar1$0] for the lowest $\nu$ 
both in $\chi$$\prime$ and $\chi$$\prime\prime$ and the peaks  
move towards a higher temperature, say, from 10 to 14 K as the $\nu$ is 
varied from 1.2 to 1200 Hz (unlike the 23.6 K transition) (Fig. 1). At this juncture, it may be stated that the neutron diffraction data on polycrystalline material\cite{10} reveal the presence of a spin-glass-like component along with the long-range ordering, which actually prompts us to attribute this frequency dependence to spin-glass-like freezing. But, in Ref. 10, both the orderings are proposed to set in at the same temperature (close to 24 K), whereas in the present study, we observe only long range ordering at 23.6 K with the spin-glass-like features appearing only at 10 K. We believe that this discrepancy arises from the differences in the degree of Pd-Si disorder  among these two forms of the material (more in the  polycrystalline form) - a finding interesting in its own right. Though this frequence dependence is characteristic of spin-glasses, it is  hard to conclude\cite{18} for such anisotropic, concentrated magnetic systems whether the transition is of a spin-glass-type. For this reason, we call this behaviour "spin-glass-like".  If these properties are truly due to spin-glass behavior, this anomaly in such a stoichiometric compound might arise due to the triangular arrangement of Tb ions (topological frustration) and/or a small degree Pd-Si disorder (see below for a discussion on crystallographic features).  In any case, if this feature in this compound is  isotropic, one should make similar observations in the data for H//[0001]. In 
sharp contrast to this expectation, we do not see (Fig. 2) any peak in 
the T range 7 - 12 K for this direction. Thus, the anisotropic nature of  the 10K-feature makes this compound an interesting one. 
However, 
the $\chi$$\prime$(T) curve  for the latter direction shows a drop at a lower 
temperature, say around 5 K, for $\nu$= 1 Hz. In other words, the former direction shows a spin-glass-like  
feature at about 10 K, whereas the latter orientation shows  similar frequency-dependent 
 $\chi$$\prime$ features around 5 K (see the inset of Fig. 1). 
These findings (low-T transitions at distinctly different temperatures for 
these two directions) are rather consistent with the features in the 
zero-field-cooled dc $\chi$ data of Ref. 12 (compare  Fig 2a and 2b of Ref. 
12). The above 5K-feature in any case can not be attributed to
"canonical spin-glass" behavior,\cite{1} as the sign of the magnetoresistance  in the vicinity of  5 K 
(see Fig. 2b of Ref. 12)  for H//[0001] is opposite (positive!) to that expected for  
spin-glasses, and that $\chi$$\prime$$\prime$(T) does not apparently exhibit a sharp feature, however weak it may be,  around 4 K.  Needless to emphasize that
the magnitude of the shift ($>$ 2 K) of 
the peak temperatures with $\nu$ for these transitions is  interestingly very large  compared to 
that known (about 1$\%$) for canonical spin glasses. It is not out of place to mention that we have also measured the isothermal magnetization (M) and we find that M varies esssentially {$\it linearly$} for H//[0001] without any hysteresis at low fields (below 20 kOe), say at 5 K, as reported in Ref. 12, whereas for H//[10$\bar1$0], M-H plot exhibits a hysteresis loop.   Therefore, if the hysteresis in M for the latter orientation arises from spin-glass-like anomaly, then clearly this behavior is  anisotropic.

{\it In order to bring out the second major point of emphasis,} we  turn to the data above 23 K. It is obvious that the $\chi$$\prime$ 
(T) exhibits a broad peak around 55 K for [0001]  as in the case of 
dc $\chi$ (T) plot (see the inset of Fig. 2, with the data taken from Ref. 12). This feature is so broad that it can not result from spin-glass freezing of some fraction of Tb ions. The present article places this on a firm footing by reporting  ac $\chi$ as a function of $\nu$, i.e., the plots for all the frequencies overlap above 25 K. 

A careful look at the crystallographic features in light of above finding is quite intriguing. As already remarked, this compound forms in a AlB$_2$-derived hexagonal crystal structure.  The ternary derivative of this structure 
presents an interesting situation (Fig. 3) crystallographically in the event\cite{19} that 
there is an ordered occupation of B-site by Pd and Si atoms in the basal plane: that is, there 
are two different chemical environments for R ions (called 2(b) and 6(h) 
sites), doubling the a-parameter.\cite{4} As a signature of this, one normally observes superstructure lines in the x-ray diffraction pattern (e.g., (101) and (110) appearing at 2$\theta$= 16.6 and 21.7 $\AA$ respectively with Cu K$_{\alpha}$ radiation). However, such superstructure lines   are usually weak (the intensity of which apparently depends upon the degree of Pd-Si disorder). Hence these   can easily escape detection. Thus, the wave-length of the neutron employed in a previous study\cite{10} on the polycrystalline form is too large  (2.44 $\AA$) that these lines could not be detectable, whereas, in the x-ray diffraction pattern recorded with a much smaller wave-length radiation (Cu K$_{\alpha}$) on the same polycrystalline sample, we could observe these lines in agreement with a previous article\cite{3} reporting doubling of a-parameter. While finalising this article, we came to know that careful neutron diffraction investigations\cite{20} on the single crystals indeed reveal the doubling of a-parameter.   The R and Pd-Si  layers are stacked alternately along 
c-direction and the arrangement of R ions in the basal plane is triangular. 
The R ions at the 6(h) site are sandwitched  by eight  Si atoms 
and four Pd atoms (with four Si and two Pd placed on the vortices of the 
hexagon on each side) along the hexagonal tube, whereas the ones at the 2(b) 
site are sandwiched by hexagons of Si in the immediate vicinity. Thus there 
are chains of R ions at the centre of the hexagonal tube made up of Pd and 
Si atoms and the R-R distances are of the order of 4 $\AA$. It has been 
established\cite{4} that the R chain at the 2(b) site can undergo magnetic 
ordering  due to finite interaction among 2(b)-2(b) chains (which are at a 
distance of about 8 $\AA$ from each other)  leaving the intervening R ions at 
the 6(h) site paramagnetic. 
We  therefore attribute the broad 55K-peak in $\chi$  to the antiferromagnetic correlations within the
chain of Tb ions 
present at the 2(b) site,  and the observed broad peak is similar to that 
predicted theoretically by Bonner and Fischer\cite{17} several years ago. In order to compare the experimental $\chi$ behavior with a model on low dimensional systems, for instance as in Ref. 21),   we have used the dc $\chi$ data (Ref. 12)  
(above 25 K) instead of ac $\chi$, as the absolute values are more reliable in the former case. We obtained a fairly good fit 
(see  continuous the line in the inset of Fig. 2 as per the equation 2 in Ref. 21) 
for an exchange interaction strength of about 35 K.    
The broad peak is however not observable for H//[10$\bar1$0] (see Fig. 1). It 
is to be remarked that the strength of this peak for H//[0001] is very low in  the 
 dc $\chi$ as well.\cite{12} Such a weak feature is  masked  
for H//[10$\bar1$0] as the $\chi$ values for the paramagnetic Tb 
ions at the 6(h) site for this orientation are large increasing at a faster rate (than the decrease of $\chi$ below 45 K in the other direction) with decreasing T.   
 It is worth noting that the paramagnetic Curie temperature is widely different (-30 and 23 K for [0001] and [10$\bar1$0] directions), unlike the situation for other analogous heavy rare-earth compounds.\cite{11,14,16} Such a strong anisotropy even in the paramagnetic state is also a characteristic of quasi-one-dimensional magnetic systems.

To conclude, Tb$_2$PdSi$_3$ apparently exhibits quasi-one-dimensional features in the 
magnetic susceptibility data (around 55 K), thereby characterizing this 
compound, to our knowledge, to be the first one among intermetallics to show this 
phenomenon. Naturally, this compound will thus serve as an ideal  model system to evolve theories to understand low-dimensional magnetism {\it dominated by indirect exchange interaction via conduction electrons} (unlike the situation till todate in insulators in which case superexchange interaction mediates magnetic coupling) and associated interchain coupling effects.  Another unusual observation is that, the 23.6K-long-range magnetic ordering   is followed by "anisotropic" 
spin-glass-like anomalies at lower temperatures; in addition, there is a large frequency dependence of the peak temperature in ac $\chi$(T) plot, uncharacteristic of canonical metallic spin-glasses.  These results clearly reveal that this compound is an exotic magnetic material.\cite{22}  

\begin{figure}
\vskip5mm
\caption{Temperature dependent  ac susceptibility ($\chi$) behavior at various frequencies for single crystalline 
Tb$_2$PdSi$_3$ for H//[10$\bar1$0]. $\chi$$\prime$ and $\chi$$\prime$$\prime$ represent real and imaginary parts respectively. The transitions  below 10 K are compared in the insets (one below the other) for two directions. The direction of the shift of the curves for T $<$ 15 K with increasing frequency (1.2, 12, 120, 1220 Hz) is shown by an horizontal arrow.}
\end{figure}
\begin{figure}
\vskip5mm
\caption{Temperature dependent  ac susceptibility ($\chi$) behavior for  various frequencies for single crystalline 
Tb$_2$PdSi$_3$ for H//[0001]. $\chi$$\prime$ and $\chi$$\prime$$\prime$ represent real and imaginary parts respectively. The peak temperature in $\chi$$\prime$ for the low temperature transition shifts to a higher temperature with increasing frequency (1.2, 12, 120 and 1220 Hz) (shown in an expanded form in the inset of Fig. 1). The $\chi$$\prime$$\prime$ is featureless in the entire temperature range with negligibly small values for lower frequencies.  The  inset shows the dc $\chi$ behavior (Ref. 12) above 25 K  with the continuous line representing a fit as described in the text.}
\end{figure}
\begin{figure}
\vskip1cm
\caption{The unit cell of Tb$_2$PdSi$_3$ in the a-b plane viewed along c-axis. Big cirles (light black at 2b site and dark black  at 6h site) represent Tb atoms, whereas smaller ones represent Pd (darker) and Si (lighter) atoms respectively.} 
\end{figure}
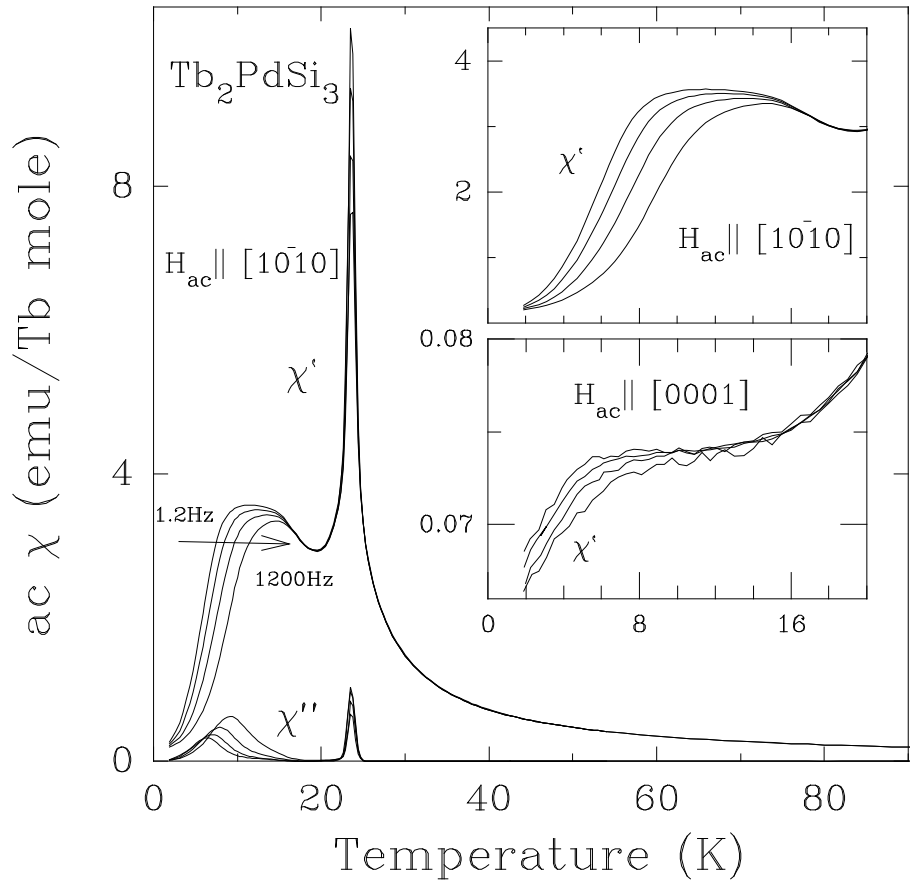
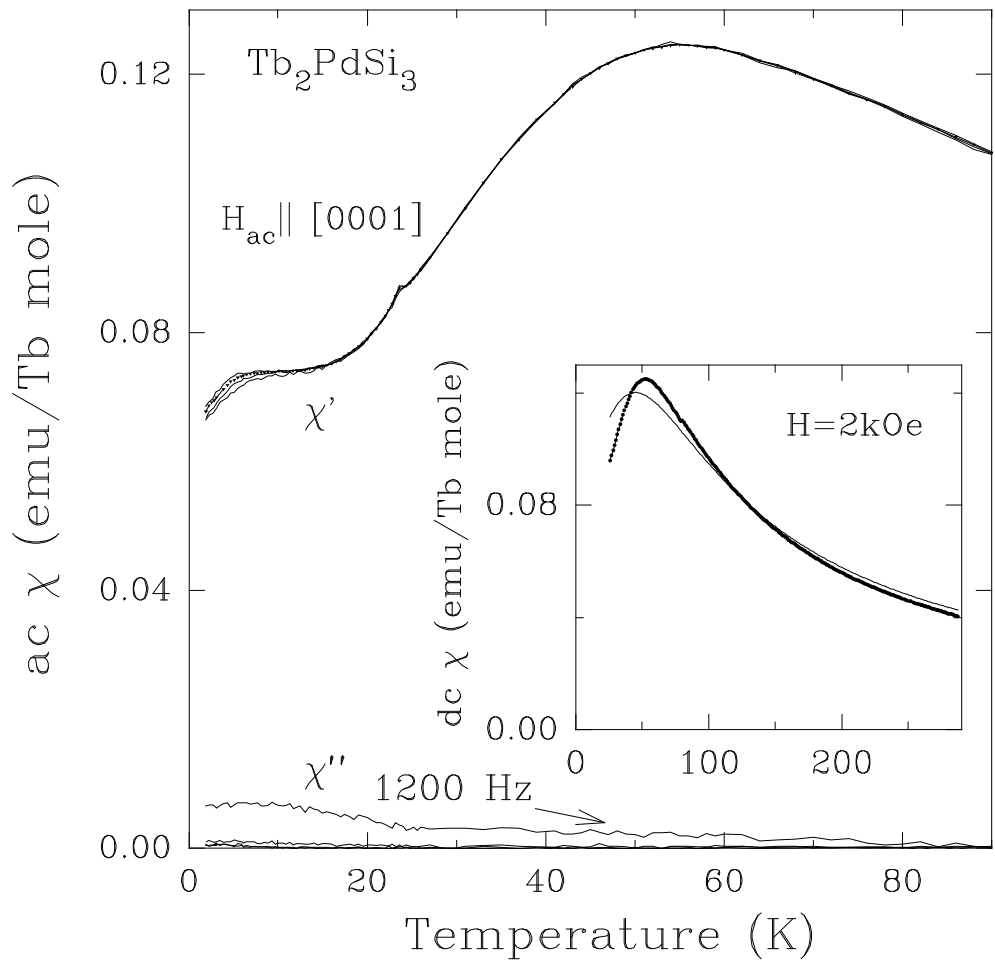
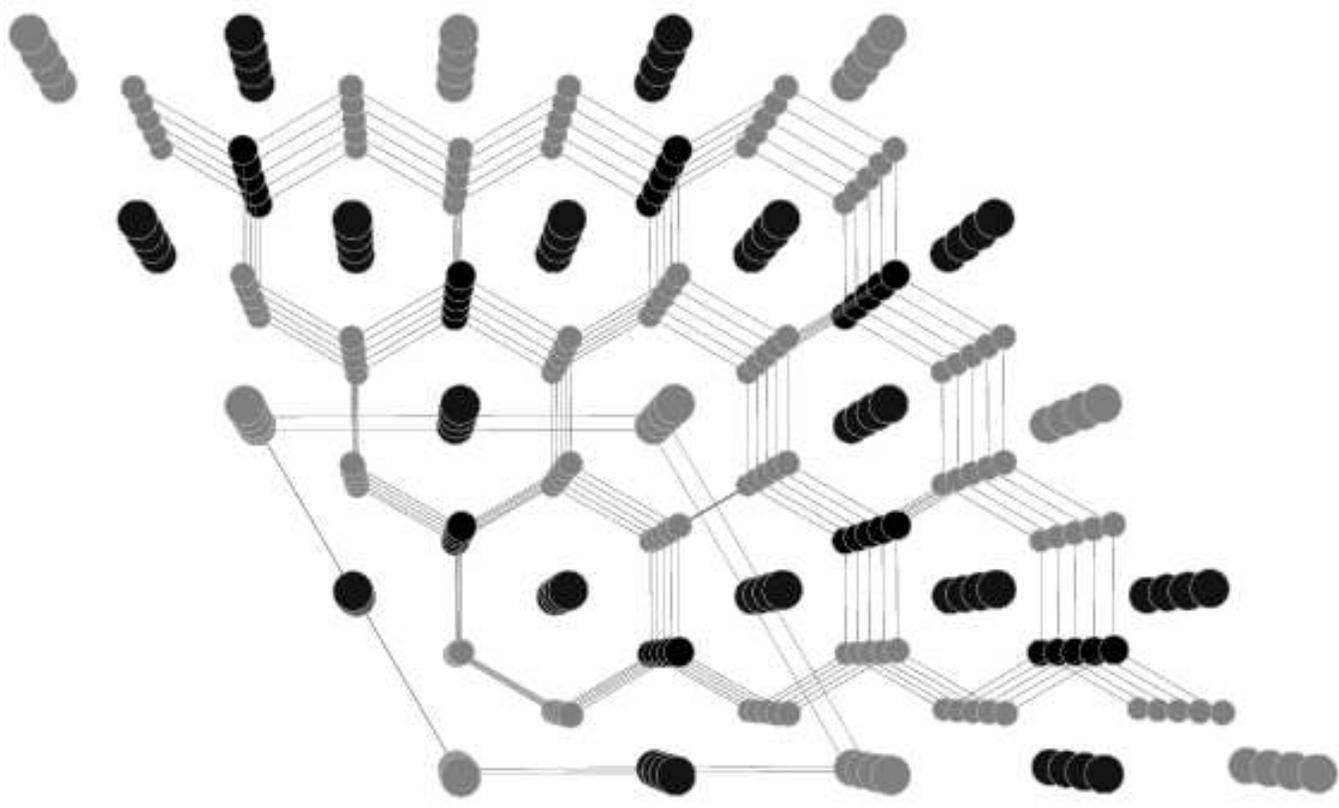
\end{document}